\documentclass[trackchanges,twocolumn]{aastex7}
\usepackage{placeins}
\usepackage{amssymb}
\usepackage{soul}
\usepackage{natbib}
\usepackage{multirow}
\usepackage{graphicx}
\usepackage{tabularx}
\usepackage{color}
\newcommand\soutm{\bgroup\markoverwith
{\textcolor{black}{\rule[0.5ex]{2pt}{0.8pt}}}\ULon}

\def\spose#1{\hbox to 0pt{#1\hss}}
\newcommand\lsim{\mathrel{\spose{\lower 3.0pt0pt\hbox{$\mathchar"218$}}
     \raise 2.0pt\hbox{$\mathchar"13C$}}}
\newcommand\gsim{\mathrel{\spose{\lower 3.0pt\hbox{$\mathchar"218$}}
     \raise 2.0pt\hbox{$\mathchar"13E$}}}
\newcommand\msun{{\rm \,M_\odot}}

\submitjournal{ApJ}

\shorttitle{Tidal features in the galaxy clusters}
\shortauthors{Chun et al.}

\graphicspath{{./}{figures/}}

\begin{document}

\title{The Role of Pre-Processing in Tidal Feature Formation within Galaxy Clusters}
\correspondingauthor{Jihye Shin}
\email{jhshin@kasi.re.kr}

\author{Kyungwon Chun}
\affil{Korea Astronomy and Space Science Institute (KASI), 776 Daedeokdae-ro, Yuseong-gu, Daejeon 34055, Korea}
\email{kwchun@kasi.re.kr}
\author{Jihye Shin}
\affil{Korea Astronomy and Space Science Institute (KASI), 776 Daedeokdae-ro, Yuseong-gu, Daejeon 34055, Korea}
\affil{University of Science and Technology (UST), Gajeong-ro, Daejeon 34113, Korea}
\email{jhshin@kasi.re.kr}
\author{Jongwan Ko}
\affil{Korea Astronomy and Space Science Institute (KASI), 776 Daedeokdae-ro, Yuseong-gu, Daejeon 34055, Korea}
\affil{University of Science and Technology (UST), Gajeong-ro, Daejeon 34113, Korea}
\email{jwko@kasi.re.kr}
\author{Rory Smith}
\affil{Departamento de F{\'i}sica, Universidad T{\'e}cnica Federico Santa Mar{\'i}a, Avenida Espa{\~n}a 1680, Valpara{\'i}so, Chile}
\affil{Millenium Nucleus for Galaxies (MINGAL)}
%1) Departamento de Fisica, Universidad Tecnica Federico Santa Maria, Avenida España 1680, Valparaíso, Chile 2) Millenium Nucleus for Galaxies (MINGAL)
\email{rorysmith274@gmail.com}
\author{So-myoung Park}
\affil{Korea Astronomy and Space Science Institute (KASI), 776 Daedeokdae-ro, Yuseong-gu, Daejeon 34055, Korea}
\email{smpark@kasi.re.kr}
\author{Songhee Nam}
\affil{Korea Astronomy and Space Science Institute (KASI), 776 Daedeokdae-ro, Yuseong-gu, Daejeon 34055, Korea}
\email{shnam@kasi.re.kr}

\begin{abstract}
We investigate the formation of tidal features, such as tidal tails, streams, and shell-like structures, composed of stars stripped from satellites within galaxy clusters.
For this, we use multiresolution cosmological $N-$body simulations with the ``galaxy replacement technique". We find that the fraction of satellites with tidal features increases with the mass of the host clusters but is not related to the dynamical state of the clusters. Although the strong tidal field in the cluster environment accelerates the mass loss of the satellites, only 20\% of tidal-featured galaxies form their tidal features purely due to tidal perturbation in the cluster environment, without any interactions with other galaxies before falling into the cluster. In contrast, the majority (80\%) is affected by the preprocessing, as they experienced merging events with other galaxies before infalling into the cluster.
Among this preprocessing population, 45\% of all tidal-featured galaxies form their tidal features after passing the pericenter of the cluster, affected by both preprocessing and the tidal field of the cluster, whereas 35\% of all tidal-featured galaxies form their tidal features before reaching the pericenter, primarily due to preprocessing.
Notably, this fraction increases from 35\% to 40-50\% when we focus only on galaxies with brighter surface brightness limits or higher stellar mass.
Therefore, our results highlight that preprocessing is an important channel for forming tidal features within clusters. However, the importance of preprocessing may be further amplified in observations, since more massive galaxies, which are commonly associated with preprocessing, are preferentially detected.

\end{abstract}


\section{Introduction}

In the concordance universe, also known as the Lambda CDM universe, galaxy mergers are the main driver not only for cosmic structure formation but also for galaxy mass growth and changes in galaxy properties \citep{press1974,fall1980,ryden1987,vbosch2002,agertz2011}. Compared to secular evolution and smooth accretion over cosmic time, the minor/major mergers between galaxies bring a radical change in star formation activity, morphology, mass growth, and other properties \citep{toomre1972,schweizer1982,barnes1991,barnes1992,mihos1994,hernquist1995,mihos1996,naab2009,vdokkum2010,newman2012,kaneko2017,ellison2018,pan2018,violino2018}. Thus, tracing merging history is crucial for understanding how galaxies have evolved in the cosmological context. 

Tidal features around the outskirts of a galaxy are remnants of recent and/or ongoing mergers between galaxies \citep{helmi1999,ferguson2002,majewski2003,toomre1972,quinn1984,barnes1988,hernquist1992,feldmann2008}. Since the tidal feature is faint and extended, their surface brightness is significantly lower than that of the galaxy's central main body, where most of the luminosity originates \citep{atkinson2013,kado-fong2018,martin2022}. Despite the difficulty in detecting them due to their low-surface brightness (LSB), the tidal features are worth observing as they can serve as evidence of the recent and ongoing galaxy merger \citep{ji2014,mancillas2019,yoon2020,huang2022}. More specifically, many analytical and numerical studies have shown that details of the tidal features such as morphological characteristics, prominence, and the number provide hints about the mass ratio of the merger, the orbital information, the time since the last merging event \citep{quinn1984,dupras1986,helmi1999,ferguson2002,majewski2003,johnston2008,lotz2008,cooper2010,sanderson2010,cooper2011,sanderson2013,ji2014,amorisco2015,hendel2015,pop2018,karademir2019,mancillas2019,khalid2024}.

Tidal features around galaxies in the LSB regime ($\mu_{\rm{V}} >$ 26.5~mag~arcsec$^{-2}$) have been unveiled by the deep imaging surveys such as SDSS stripe 82 \citep{kaviraj2010,jiang2014,peters2017}, MATLAS \citep{duc2015,bilek2020}, NGVS \citep{ferrarese2012}, Dragonfly Nearby Galaxies Survey \citep{merritt2016}. See Table 1 of \cite{bilek2020} for more references. Across the observations, statistics on the fractions of galaxies with tidal features are quite inconsistent, ranging from 0.06 to 0.7, due to differences in the galaxy sample, the image depth, and the identification criteria \citep{bilek2020}. Using four different cosmological hydrodynamic simulations of New Horizon \citep{dubois2021}, EAGLE \citep{crain2015,schaye2015}, IllustrisTNG \citep{marinacci2018,naiman2018,nelson2018,springel2018}, and MAGNETICUM \citep{teklu2015}, \cite{khalid2024} created the LSST-like mock images for galaxies by a consistent sample selection criteria and a uniform methodology. They found that the tidal feature fraction ($f_{tide}$) is quite similar across the simulations, ranging from  0.32 to 0.4. It suggests that the variation of $f_{tide}$ observed in different surveys may be due to inconsistent methodology, especially due to different image depths \citep{martin2022}. Furthermore, their result implies that the direct driver for generating the tidal feature is not baryonic physics but rather gravitational forces, which is consistent among the simulations \citep{khalid2024}.   

Galaxy clusters are the most massive objects gravitationally bound, and the place where galaxies are being disassembled vigorously. Therefore, cluster galaxies have been key research topics for studying environmental effects on galaxy characteristics \citep{richstone1976,gnedin2003b,Wetzel2010}. Compared to the field and group environments, the cluster galaxies are known to have the smaller $f_{\text{tide}}$ value, although its significance is low due to limited sample size \citep{tal2009}. To increase the statistics, \cite{adams2012} analyzed 54 different galaxy clusters of 0.04 $<$ z $<$ 0.15 containing 3551 early-type galaxies and identified tidal features. They found that $\sim3\%$ of cluster early-type galaxies have tidal features of $\mu <$ 26.5~mag~arcsec$^{-2}$, and that there is no evidence of dependency of $f_{\text{tide}}$ with the local density but the decreasing trend for the smaller clustercentric radius than $0.5R_{vir}$. From the deficit of tidal-featured galaxies in the inner region, we infer that merging is a rarer event in the inner region because of their higher velocity \citep{binney1987,gnedin2003a} and that the tidal features are more vulnerable to being erased in the inner region \citep{rudick2009}. However, there is another possibility that the contaminated background brightness coming from the brightest cluster galaxy (BCG) and intra-cluster light (ICL) may affect the identification of tidal features at the inner cluster region since the background brightness is a function of the clustercentric radius. Furthermore, if one considers interaction with the BCG as galaxy mergers, then the merging frequency for the inner cluster region is not as low as we had expected. 

On the other hand, a considerable portion of cluster galaxies enters the cluster as a member of galaxy groups or through filaments and thus preprocessed before entering the cluster environment \citep{fujita2004,mihos2004,sheen2012,villalobos2012,wetzel2013,Yi2013,villalobos2014,joshi2017,han2018,oh2018,benavides2020,haggar2023}. The fraction of the preprocessed galaxy is estimated to range from 15\% to 50\% based on N-body simulations and semi-analytic models \citep{berrier2009,mcgee2009,joshi2017}. Since the preprocessed galaxies have lost significant mass before entering the cluster, their stellar components are prone to be tidally stripped compared to other galaxies that first infall into the cluster \citep{smith2015,han2018}. 
In particular, \cite{han2018} showed that among the cluster galaxies at $z=0$ that have suffered a significant mass loss (above 80\% of their peak mass), 74\% were previously members of another host before entering the cluster. This indicates the possibility that preprocessing plays an important role in forming the tidal features within clusters.
A compelling example of this is provided by \cite{roman2023}, who discovered a giant thin stellar stream in the Coma galaxy cluster, located far from the cluster center at $D\sim0.8$Mpc.
They suggested that the thin stream may originate from the preprocessed dwarf galaxy by comparing with an analog found in Illustris TNG-50.

In this study, we aim to trace how tidal features of galaxies form and evolve in and around cluster environments. For this, we perform cosmological N-body simulations targeting 84 different clusters using the ``galaxy replacement technique" \citep[GRT;][]{chun2022,chun2023,chun2024}, which is designed and optimized for describing the gravitational evolution of stellar components in a hierarchical merging context. To understand the effect of preprocessing on tidal features, we trace the merger history of individual galaxies that have ever entered the cluster region. In this study, the tidal feature is identified using an escape velocity profile to avoid dependence on the surface brightness (SB) limits.

This paper is organized as follows. In Section \ref{method}, we introduce simulations using the GRT and describe methodologies for identifying galaxy and tidal features. In Section \ref{result}, we explore several factors affecting the tidal feature formation of cluster galaxies and how the fraction of tidal-featured galaxies is dependent on the projected clustercentric distance, the galaxy's stellar mass, and the SB limits. We discuss and summarize the results in Section \ref{discussion} and \ref{sec:summary}, respectively.

\section{Method}\label{method}

\subsection{Simulations using Galaxy Replacement Technique}

In this study, we examine the tidal features of cluster galaxies using cosmological multi-resolution N-body simulations targeting 84 different clusters, which are named the ``GRT clusters" \citep{chun2023,chun2024}. The GRT enables us to efficiently trace the impact of gravitational tides in a fully cosmological context without expensive hydrodynamics \citep{chun2022}. Briefly, the ``GRT clusters" run is conducted as follows: 1) low-resolution uniform box DM-only simulations are performed first, 2) target clusters are selected and defined as the GRT clusters, 3) low-resolution DM halos at z$>0$ that will later constitute the growth of the GRT clusters are selected, 4) high-resolution model galaxies are generated to be substituted for the low-resolution DM halos either (i) when they reach their peak mass $M_{\text{peak}}$, which is the maximum mass of the DM halos before first falling into a more massive halo, or (ii) when they first accrete a replaced satellite, and 5) the multi-resolution resimulations are performed to trace the evolution of the GRT clusters until z = 0. 

\begin{figure}
    \centering
    \includegraphics[width=0.48\textwidth]{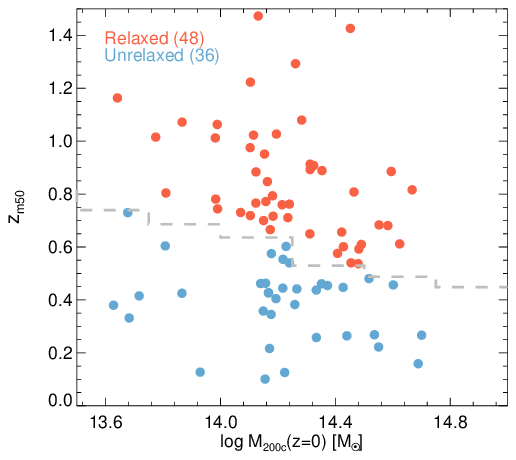}
    \caption{The relation between M$_{\rm{200c}}$ and $z_{\rm{m50}}$ of the GRT clusters at $z=0$. The filled red and blue circles indicate relaxed and unrelaxed clusters. The dashed gray line is the median $z_{\rm{m50}}$ of all clusters in the N-cluster run within a logarithmic mass bin of 0.25 dex.}
    \label{fig:zm50}
\end{figure}

\begin{figure*}
\centering
\includegraphics[width=0.95\textwidth]{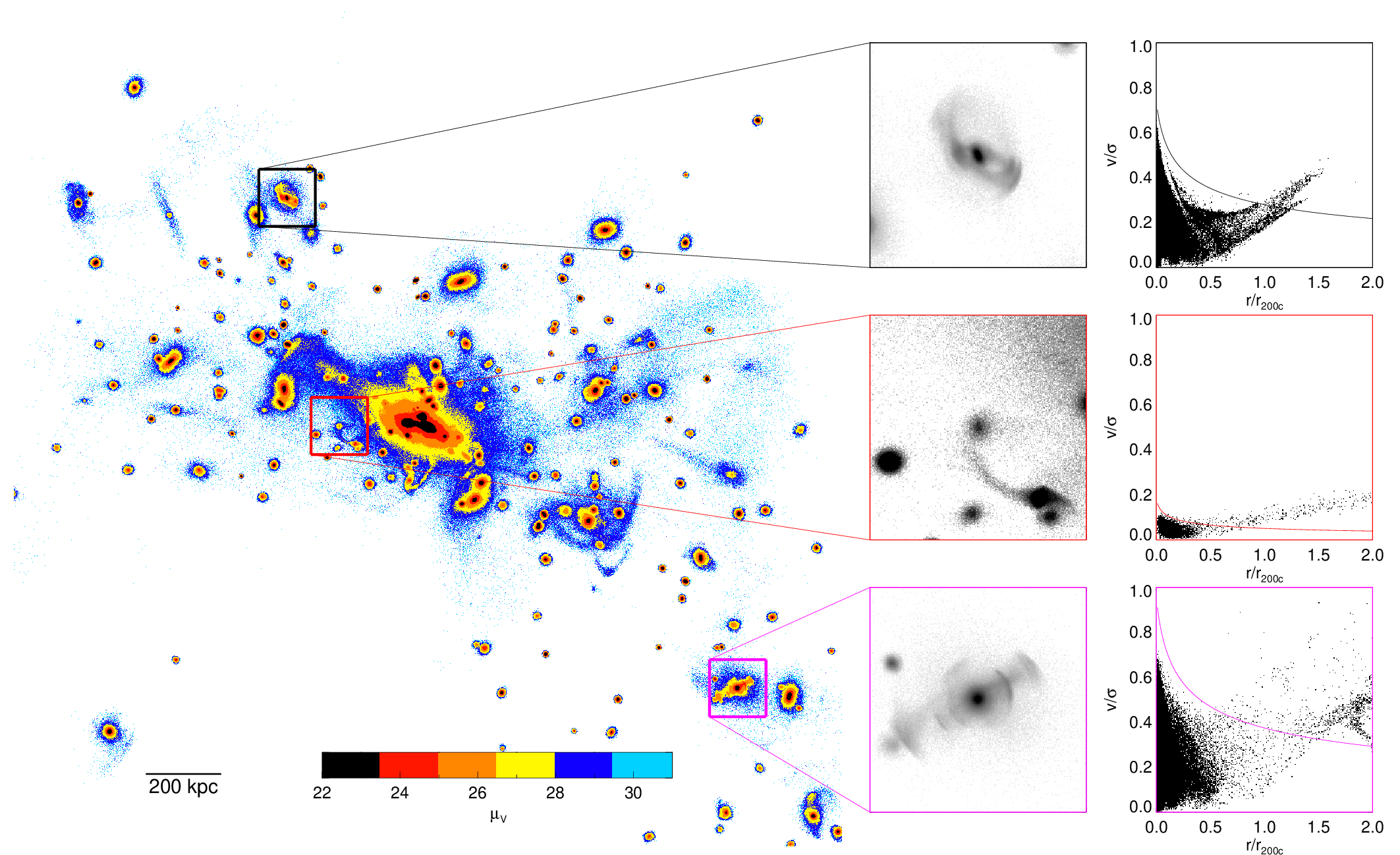}
\caption{The V-band surface brightness ($\mu_V$) map of the most massive GRT cluster at $z=0$. Color bars indicate the V-band surface brightness. The right columns show the stellar distribution and phase-space diagram of the representative tidal tail, stream, and shell-like structures from the upper panels. In each phase-space diagram, the solid lines indicate the escape velocities ($v_{\rm{esc}}$) derived using the current members, and the stars with velocities greater than $v_{\rm{esc}}$ are considered to be in the unbound regime.}
\label{fig:map}
\end{figure*}

The low-resolution simulation, named the $N-$cluster run, is a set of 64 uniform boxes of (120 Mpc/h)$^3$, whose DM particle mass is $10^9\msun$/h and the gravitational softening length is fixed at 2.3 kpc/h on a comoving scale. This simulation set is used in many works \citep{smith2022a,smith2022b,chun2022,jhee2022,kim2022,yoo2022,awad2023,chun2023,dong2024,kim2024,chun2024}. The high-resolution galaxy models include a DM halo and a stellar disk, with particle mass of $5.4 \times 10^6 \msun$/h for the DM and $5.4 \times 10^4\msun$/h for the stars, and gravitational softening lengths of approximately 100~pc/h and 10~pc/h, respectively. Here, the DM halos replaced as model galaxies are limited to be more massive than $10^{11}\msun$/h ($N_{DM}$ = 100) when they reach $M_{\text{peak}}$. The stellar mass for the high-resolution galaxy model is decided based on the time-evolving halo abundance matching \citep{behroozi2013a}. Among halos at $z=0$, a total of 84 clusters in a mass range of $13.6 < \log M_{\text{200c}}/\msun < 14.8$ are selected in a way that their $z_{\text{m50}}$ values are evenly distributed, where $M_{\text{200c}}$ is the total mass within a radius ($R_{\text{200c}}$) where density drops to 200 times of the critical density of the Universe, and $z_{\text{m50}}$ is an epoch when a halo first accrete half of their final $M_{200c}$ (see Fig. \ref{fig:zm50}). 
We classify the clusters with $z_{\rm{m50}}$ above the median value of all clusters in the $N-$cluster run as relaxed clusters (the filled red circles) and others as unrelaxed clusters (the filled blue circles). This panel shows the decrease of $z_{\rm{m50}}$ with increasing cluster mass due to the hierarchical mass growth in $\Lambda$CDM cosmology.
More details of the ``GRT'' and the ``GRT clusters'' are available in \cite{chun2022,chun2023,chun2024}.

In the GRT simulation, as we do not replace the low-mass halos (M$_{\text{peak}} < 10^{11}\msun$/h) with the high-resolution galaxy models, the population of low-mass satellites in the cluster is not complete.
Therefore, in this work, we use the satellites with M$_{\rm{gal}} > 10^9\msun$, where M$_{\rm{gal}}$ is the galaxy stellar mass of the satellites.
We find the mass function of satellites with M$_{\rm{gal}} > 10^9\msun$ in the clusters is similar to that in the clusters in other cosmological hydrodynamic simulations (e.g., Illustris TNG-50, TNG-100).

\subsection{Identification of galaxies and their tidal features}

\begin{deluxetable*}{ll}
\caption{Definitions of time-related parameters used in this section}
\label{tab:properties}
\tablewidth{0pt}
\tablehead{
\colhead{Parameters} & \colhead{Definition} 
}
\startdata
$t_{\text{form}}$ & The lookback time when the galaxy is formed\\
$t_{\text{infall}}$ & The lookback time when the galaxy first falls into the cluster\\
$t_{\text{tide}}$ & The lookback time when the galaxy forms tidal features\\
$t_{\text{peak}}$ & The lookback time when the galaxy has its maximum mass before falling into a more massive galaxy,\\
 & or when the galaxy first accretes a replaced satellite\\
$\Delta t_{\text{tide}}$ & The time elapsed from the galaxy's infall to tidal feature formation, i.e., $\Delta t_{\text{tide}} = t_{\text{infall}} - t_{\text{tide}}$\\
$\Delta t_{\text{pre}}$ & The time interval during the galaxy has interacted with other galaxies before entering the cluster
\enddata
\end{deluxetable*}

Structures such as clusters, galaxies, and satellites are identified with the modified 6D phase space halo finder ROCKSTAR \citep{behroozi2013b}, and the merger tree is built using Consistent Trees \citep{behroozi2013c}. We define the galaxies that have ever entered $R_{200c}$ of the cluster as cluster galaxies. When the cluster galaxy first crosses the $R_{200c}$ of the cluster, all its member particles are tagged as `infall members' of the galaxy and traced until $z=0$ to be used for identifying tidal features.
The infall members with velocities greater than escape velocity ($v_{\text{esc}}$) at a given distance from the galaxy center are considered to be in unbound regime, while those with velocities less than $v_{\text{esc}}$ are considered to be in galaxy regime, which is bound to the galaxy. Here, we derive $v_{\text{esc}}$ profile using the `current members', defined as the galaxy members identified at a given time. Central and satellite galaxies are analyzed separately. 

From the moment when the infall members in the unbound regime are more than $10^7\msun$ ($N_{\text{star}}\gsim127$), the galaxy is classified as a tidal-featured galaxy. This mass corresponds to the 1\% of the mass criteria used for our galaxy sample. 
In the GRT simulations, all replaced galaxy models are more massive than $10^7\msun$ when they are inserted into the DM halos. Therefore, this threshold ensures that tidal features from the disruption of all well-resolved galaxies, including those produced by the complete disruption of relatively low-mass satellites, are captured.
To clearly compare galaxies with the tidal feature to those without it, galaxies whose infall members in the unbound regime are less than $10^6\msun$ are classified as non-tidal galaxies, and the intermediate galaxies between $10^6\msun$ and $10^7\msun$ are excluded in the comparison analysis. Classifying the tidal-featured galaxy with different $N_{\text{star}}$ thresholds shows that our results in Section \ref{result} are not sensitive to the threshold. Note that our classification using $v_{\text{esc}}$ for the tidal features focuses on the internal dynamics of each galaxy rather than its morphological features identified by visual inspection. We find that two different methods especially make a difference in the classification of tidal feature in low-mass galaxies (M$_{\rm{gal}}~<~10^{10}\msun$).
We discuss these differences in Section \ref{discussion}.

Figure \ref{fig:map} shows a V-band surface brightness map of stars located in a 2~Mpc cube centered on the most massive GRT cluster. Three squares indicate representatives of the tidal-featured galaxies identified using our classification criteria. The intra-cluster region is filled with stars in the LSB regime of $\mu_{V} >$ 26.5~mag~arcsec$^2$, while finer structures originate from tidal features, such as tail, stream, and shell.
The right panels show the stellar distribution near tidal-featured galaxies and the phase-space diagram of their infall members. In each phase-space diagram panel, we can see that some infall members have velocities greater than $v_{\text{esc}}$ indicated by the solid line. These stellar particles are considered to be in the unbound regime, and in Section \ref{result}, we investigate how and when they become unbound from the galaxy and contribute to the formation of tidal features.
Note that, due to our definition, the tidal features in this study only include the tidal stream, tail, and shell-like structures, but do not include bound tidal features (tidal-induced bar, strong spiral, asymmetric halo, double nucleus, etc.).
Ram pressure stripping \citep{gunn1972} can also induce morphological transformations in galaxies, which may in some cases be difficult to distinguish from structures formed by tidal interactions \citep{smith2025}. However, these typically manifest as features in the bound stellar disk rather than as unbound tidal features. Therefore, although ram pressure stripping is an important process that affects the evolution of cluster galaxies, it is unlikely to significantly alter the main conclusions of our study.

\begin{figure}[hb!]
    \centering
    \includegraphics[width=0.48\textwidth]{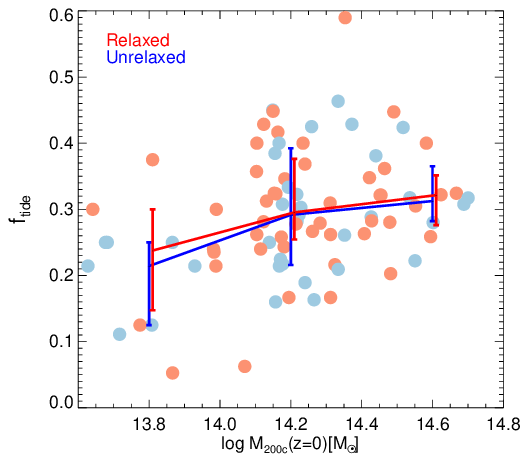}
    \caption{The relation between M$_{\rm{200c}}$ and $f_{\rm{tide}}$ of the GRT clusters at $z=0$. The filled red and blue circles indicate the relaxed and unrelaxed clusters and the solid lines indicate the median value of $f_{\rm{tide}}$ of the GRT clusters ($13.6 < \log$ M$_{\rm{200c}} < 14.8$) within the three host mass bins with 0.4~dex. The error bar caps indicate the first and third quartiles of the $f_{\rm{tide}}$ at each mass bin.}
    \label{fig:ftide}
\end{figure}

\section{Result}\label{result}

\begin{figure*}[ht!]
\centering
\includegraphics[width=0.9\textwidth]{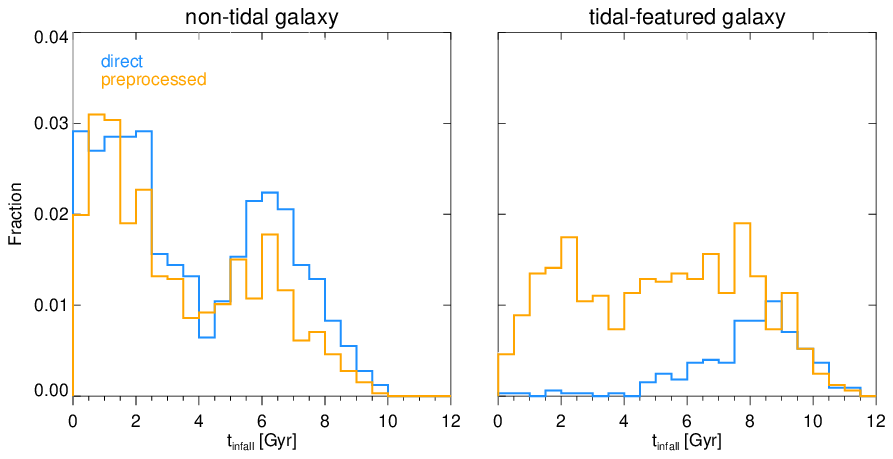}
\caption{Histograms of infall lookback time(t$_{\rm{infall}}$) of non-tidal and tidal-featured galaxies. The blue and yellow histograms indicate the direct and preprocessed infallers, respectively. The fraction indicates the number of direct or preprocessed galaxies among all galaxies with M$_{\rm{gal}} > 10^9 \msun$ within the 0.5~Gyr bin.}
\label{fig:type_infall}
\end{figure*}

In this section, we investigate how many satellite galaxies have tidal features at $z=0$, as well as how and when the features are generated in the cluster environment.
For this, we use various time-related parameters of satellite galaxies, such as the time of tidal feature formation, infall time, etc.. For clarity, we summarize their definitions in Table~\ref{tab:properties}. 
Additionally, we refer to the halo mass of satellite galaxies at each time $t_{\rm x}$ as $M_{\rm x}$ (e.g., $M_{\rm{form}}$ at $t_{\rm{form}}$).
Figure \ref{fig:ftide} shows the $f_{\text{tide}}$ value for the 84 GRT clusters at $z=0$.
They are highly scattered, ranging from 0.05 to 0.6, but we do not find a significant dependency of $f_{\text{tide}}$ on the dynamical states of clusters. On the other hand, $f_{\text{tide}}$ of the clusters increases with the mass of clusters. The median values are from 0.21 to 0.32 for the three host mass bins of $13.6 < \log$ M$_{\rm{200c}} < 14$, $14.0 < \log$ M$_{\rm{200c}} < 14.4$, and $14.4 < \log$ M$_{\rm{200c}} < 14.8$ at $z=0$.

\subsection{Factors affecting the tidal feature formation of cluster galaxies}

Among the 3,262 cluster galaxies with M$_{\rm{gal}} > 10^9\msun$, 985 are classified as tidal-featured galaxies and 1,897 as non-tidal galaxies. The tidal-featured and non-tidal galaxies are divided into direct and preprocessed infallers based on whether they have interacted with other luminous galaxies before the infall. 
Here, we define a luminous galaxy as one with a stellar mass greater than $10^7~\msun$. In the GRT simulations, all galaxies exceed this limit when they are inserted into the DM halos. This ensures that preprocessing events occur between well-resolved galaxies, whereas direct infallers interact only with less massive or unresolved halos before falling into the cluster.
As we mentioned above, we exclude the intermediate galaxies in the analysis.
Total number of direct and preprocessed infallers is 1,276 and 1,606, respectively. Among the direct infallers, galaxies classified as tidal-featured and non-tidal galaxies are 209 (16\%) and 1,067 (84\%), respectively. Within the preprocessed infallers, the percentage of the tidal-featured galaxies increases to 776 (48\%), while that of non-tidal decreases to 830 (52\%). The difference in percentages originates from the fact that part of the DM halo of preprocessed infallers has been stripped before their infall, and thus, star particles are more susceptible to be stripped \citep{smith2016}. Indeed, the typical infall mass of the preprocessed infallers is 76\% of its peak mass (M$_{\text{peak}}$), while that of direct infallers is 96\%.

\begin{figure}[ht!]
\centering
\includegraphics[width=0.45\textwidth]{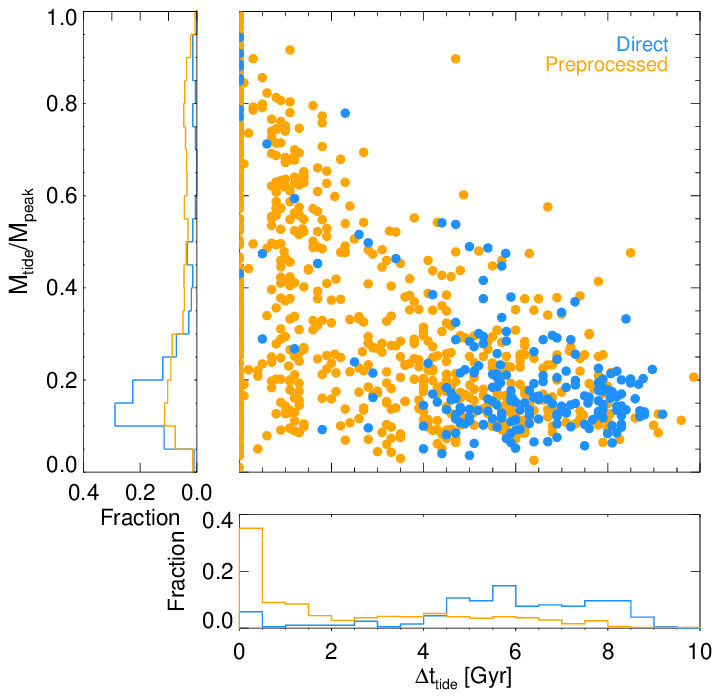}
\caption{Scatter plot and histograms of the halo mass ratio $M_{\text{tide}}/M_{\text{peak}}$ and the time taken to form the tidal features after cluster infall, $\Delta t_{\rm{tide}}$. Here, $M_{\text{tide}}$ and $M_{\text{peak}}$ are the halo masses at the time of tidal feature formation ($t_{\rm{tide}}$) and at the time when the halo has peak mass ($t_{\rm{peak}}$), respectively.
The filled blue and yellow circles indicate the direct and preprocessed infallers. 
The cluster galaxies that form their tidal features before falling into the cluster are placed at $\Delta t_{\rm{tide}}=0$ for convenience.}
\label{fig:mpeak}
\end{figure}

Figure \ref{fig:type_infall} shows $t_{\text{infall}}$ distributions of direct and preprocessed infallers for the non-tidal and tidal-featured galaxies. % \tcm{, where $t_{\rm{infall}}$ is the lookback time that the cluster galaxy first falls into the cluster}.
The $t_{\text{infall}}$ distribution of the non-tidal galaxies shows similar two peaks for direct and preprocessed infallers, around 1.0 and 6.0~Gyr.
We find that the two peaks are related to the dynamical states of clusters. Specifically, since cluster galaxies typically fall into relaxed clusters earlier than unrelaxed clusters, the earlier peak is primarily driven by satellites in relaxed clusters, and the later peak is associated with satellites in unrelaxed clusters.
On the other hand, that of the tidal-featured galaxies shows a significant difference.
Most direct infallers classified as tidal-featured galaxies entered the cluster more than 4~Gyr ago, with a peak around 9.0~Gyr. 
In contrast, preprocessed infallers can be classified as tidal-featured galaxies even if their infall occurred within the last 4~Gyr. The difference originates from the fact that the direct infallers take more time to be classified as the tidal-featured galaxy after entering the cluster ($\Delta t_{\text{tide}}$) compared to preprocessed infallers (see bottom panel of Fig. \ref{fig:mpeak}).

Direct infallers take approximately $6\pm2$~Gyr to lose a significant amount ($>~ 80\%$ of $M_{\rm{peak}}$) of mass and thereby become classified as tidal-featured galaxies. Therefore, direct infallers that have entered the cluster within the recent 4~Gyr are less likely to be classified as tidal-featured galaxies. In contrast, the preprocessed infallers that entered within the recent 4~Gyr can be classified as tidal-featured galaxies because they have lost a significant amount of DM halos before the infall and thus take the shorter $\Delta t_{\text{tide}}$ compared to that of the direct infallers. For the same reason, the $t_{\text{infall}}$ distribution of the preprocessed infallers extends to more recently than the direct infallers.
However, if we consider $t_{\text{infall}}$ of preprocessed infallers that were satellites of more massive hosts as the time they first fell into those hosts before falling into the cluster, their $t_{\text{infall}}$ distribution shows a similar trend to that of direct infallers, with a peak around 8.5 Gyr.

Figure \ref{fig:mpeak} shows a scatter plot of tidal-featured galaxies against $\Delta t_{\text{tide}}$ and the mass ratio ${M_{\text{tide}}}/{M_{\text{peak}}}$, where $M_{\rm{tide}}$ is the halo mass of the galaxies at $t_{\rm{tide}}$. 
The $\Delta t_{\text{tide}}$ distribution of the preprocessed infallers is biased to the lower value compared to that of the direct infallers, indicating the preprocessed infallers are classified as tidal-featured galaxies in a shorter time scale. Moreover, among the preprocessed infallers with tidal features, 43\% form their tidal features before passing the first pericenter of the cluster. Specifically, 9\% form the tidal features after entering the cluster but before reaching the pericenter, while 34\% already form their tidal features before the entry.

At the moment when the cluster galaxies start to form the tidal feature, they typically lose more than 80\% of their peak mass (${M_{\text{tide}}}/{M_{\text{peak}}} < 0.2$), which is in line with results from a hydrodynamic simulation \citep{smith2016} showing that dark matter is stripped more significantly before stellar stripping. 
Meanwhile, some tidal-featured galaxies retain a relatively more significant fraction of their peak mass. Specifically, 30\% of the total tidal-featured galaxies have ${M_{\text{tide}}}/{M_{\text{peak}}}$ value greater than 0.5, and most of them (89\%) are preprocessed infallers that have or had have satellites before the cluster entry. 
In these cases, the tidal features do not originate from the main stellar body of the central galaxy, but instead from stars that were originally bound to its satellites and subsequently stripped or disrupted.
%Thus, the tidal features of the high ${M_{\text{tide}}}/{M_{\text{peak}}}$ galaxies originate not from their main stellar body but from satellites that had been stripped and/or disrupted.

To understand how tidal perturbation in the cluster affects the formation of the tidal feature of the cluster galaxy, we quantify the impact of the tidal perturbation during the orbit after falling into the cluster\footnote{In the cluster environment, the tidal features are also formed by the merging events between cluster galaxies. However, as a relative velocity between the cluster galaxies is too high, the clusters and massive groups are the worst environment for merging events \citep{ostriker1980,oh2018,pearson2024}. Therefore, we only consider the tidal perturbation by the potential of the host cluster for the formation of the tidal features.}. The perturbation parameter $P$ can be formulated as $(R^2/GM_g)(GM_cR/d^3)$, where $M_c$ is the enclosed cluster mass inside the clustercentric radius $d$, $R$ is the effective radius of the galaxy, and $M_g$ is half of the total stellar mass \citep{byrd1992,lee2018}. The $P$ value corresponds to the mean ratio between the tidal perturbation by the cluster and the gravitational acceleration of the galaxy by itself at $R$ after falling to the cluster. During the galaxy orbit in the cluster, the $P$ value is averaged as follows: 

\begin{equation}
    \log P = \log \frac{1}{\Delta t} \int_{t_0}^{t_1} \frac{R^2(t)}{M_g(t)} \frac{M_c(t)R(t)}{d^3(t)} \,dt, 
\end{equation}
where $\Delta t=t_1-t_0$. Here, $t_1$ is $t_{\rm{infall}}$ for both the tidal-featured galaxies and non-tidal galaxies, while $t_0$ is $t_{\rm{tide}}$ in the case of tidal-featured galaxies and $z=0$ in the case of the non-tidal galaxies.

\begin{figure}
\centering
\includegraphics[width=0.45\textwidth]{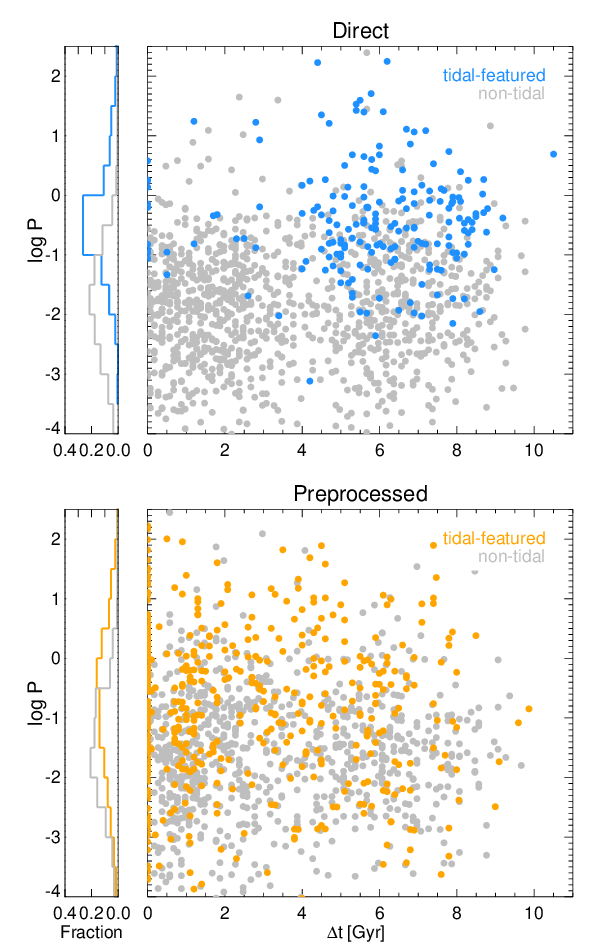}
\caption{The impact of tidal perturbation on the cluster galaxies. The upper and bottom scatter plots indicate the degree of tidal perturbation ($\log P$) that direct and preprocessed infallers experience in the cluster during $\Delta t$. Here, $\Delta t$ is the time interval between the infall lookback time ($t_{\rm{infall}}$) and the lookback time when the tidal feature forms ($t_{\rm{tide}}$) in the case of tidal-featured galaxies (filled colored circles) and z=0 in the case of the non-tidal galaxies (filled gray circles). The left histograms show the distribution of $\log P$.}
\label{fig:logp}
\end{figure}

The upper panel of Figure \ref{fig:logp}, which presents the results for the direct infallers, shows that the $\log P$ distribution of the tidal-featured galaxies has a higher peak value at $\sim$ -0.5 compared to the non-tidal galaxies, which have a peak at $\sim$-2. When $\Delta t$ is the same, the tidal-featured galaxies have a higher $\log P$ value compared to the non-tidal galaxies. This implies that $\log P$ is one of the basic factors determining the formation of the tidal features.
The $\log P$ distributions of tidal-featured and non-tidal galaxies for the preprocessed infallers are shown in the bottom panel of Figure \ref{fig:logp}. The difference between the $\log P$ distributions of the tidal-featured and non-tidal galaxies decreases compared to that of the direct infallers. The reduced difference of the $\log P$ distribution between the tidal-featured and non-tidal galaxies suggests that the tidal features of the preprocessed galaxies are less directly related to the tidal perturbation inside the cluster region. On the other hand, the formation of their tidal features is related to the degree of the preprocessing before falling into the cluster.

\begin{figure*}
\centering
\includegraphics[width=0.9\textwidth]{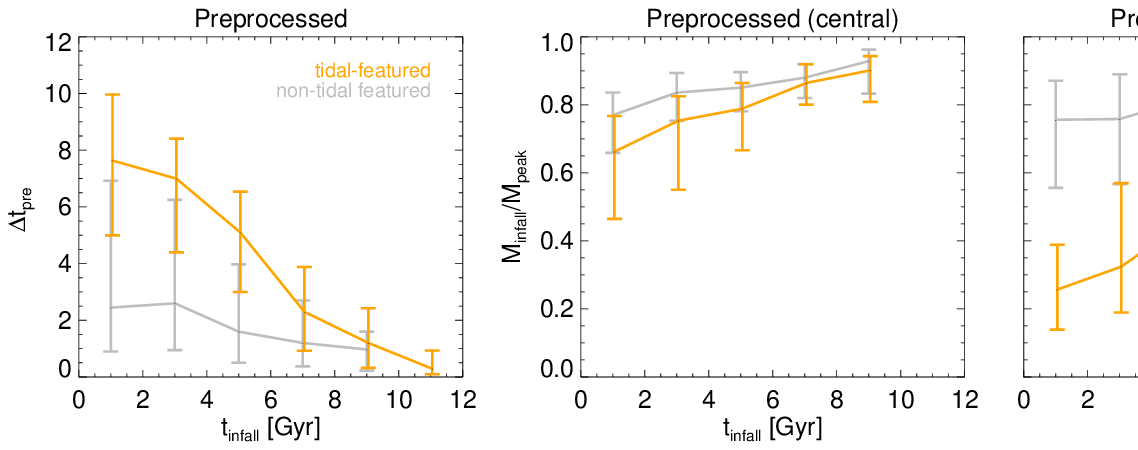}
\caption{The degree of preprocessing of the preprocessed infallers. The left panel shows the relation between the infall lookback time ($t_{\rm{infall}}$) and the time interval during a preprocessed infaller has interacted with other galaxies before entering the cluster ($\Delta t_{\rm{pre}}$). The middle and right panels indicate the mass ratio M$_{\text{infall}}/$M$_{\text{peak}}$ of preprocessed infallers, whether they are central or satellite galaxies before falling into the cluster. The yellow and gray solid lines indicate the median value of $\Delta t_{\rm{pre}}$ and M$_{\text{infall}}/$M$_{\text{peak}}$ within 2~Gyr time bin. Each error bar shows the first and third quartiles of each value.}
\label{fig:pre}
\end{figure*}

We now measure the time interval $\Delta t_{\text{pre}}$ during a preprocessed infaller has interacted with other galaxies before entering the cluster, and the mass ratio M$_{\text{infall}}/$M$_{\text{peak}}$.%, where M$_{\text{infall}}$ is M$_{\rm{200c}}$ at the cluster entry. 
Figure \ref{fig:pre} shows $\Delta$ $t_{\text{pre}}$ and M$_{\text{infall}}/$M$_{\text{peak}}$ as a function of $t_{\text{infall}}$ for the preprocessed galaxies. The $\Delta$ $t_{\text{pre}}$ value of the tidal-featured galaxies is notably higher than that of the non-tidal galaxies (left-hand panel of Figure \ref{fig:pre}), indicating that tidal-featured galaxies have interacted with other galaxies longer. The outcome of the longer interaction appears as the small M$_{\text{infall}}/$M$_{\text{peak}}$ (middle and right-hand panels of Figure \ref{fig:pre}). 
Compared to the non-tidal galaxies, the tidal-featured galaxies show significantly smaller M$_{\text{infall}}/$M$_{\text{peak}}$ in case of entering the cluster as satellites of other central galaxies (right-hand panel), while that of central galaxies is relatively insignificant (middle panel). As a result, recent infallers preprocessed as satellites of other massive hosts tend to have shorter formation time ($\Delta t_{\text{tide}}$) of their tidal features.
However, if we measure the formation time of tidal features for infallers preprocessed as satellites of other massive hosts from the time they first fell into those hosts (i.e., $\Delta t_{\text{tide}} + \Delta t_{\text{pre}}$), the timescale is similar to the $\Delta t_{\text{tide}}$ of direct infallers, with a peak around 5.6 Gyr. This indicates that the mass loss of cluster galaxies due to preprocessing is as significant as that in the cluster environment \citep[e.g.,][]{han2018}.

\subsection{Fraction of the tidal-featured galaxies with observational limitations}\label{limit}

\begin{figure*}
\centering
\includegraphics[width=0.9\textwidth]{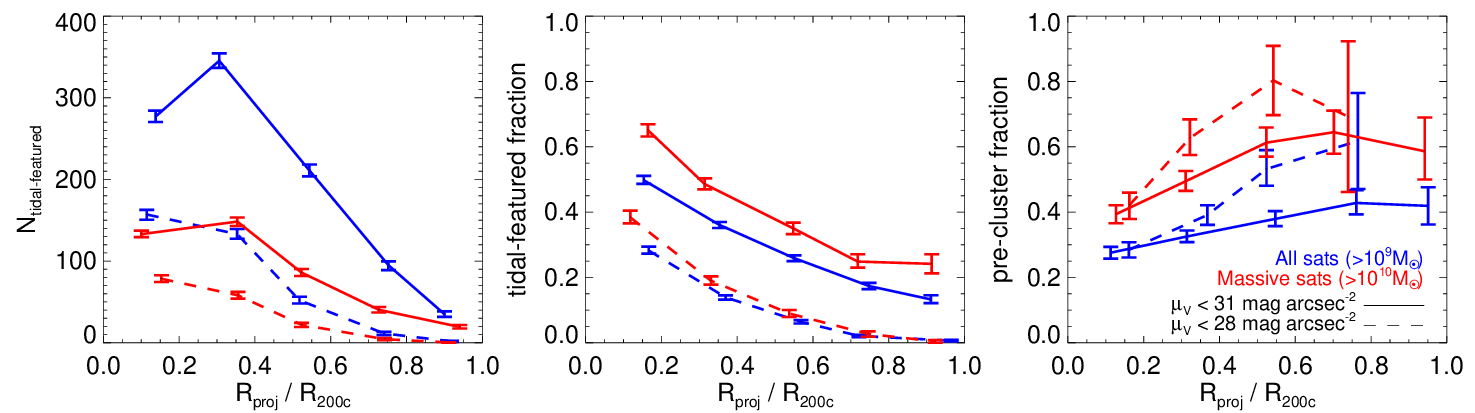}
\caption{The radial profile of tidal-featured galaxies in the cluster. From the left, each panel shows the projected radial profile of the number of tidal-featured galaxies, the tidal-featured fraction, and the pre-cluster fraction. In each panel, the blue and red solid lines are for all tidal-featured galaxies and massive tidal-featured galaxies ($M_{\rm{gal}}~>~10^{10} \msun$) brighter than 31~mag~arcsec$^{-2}$ (the solid line) and 28~mag~arcsec$^{-2}$ (the dashed line). The error bars represent the 1$\sigma$ uncertainties derived from bootstrap sampling. The bins containing fewer than five galaxies are omitted.}
\label{fig:rprof}
\end{figure*}

\begin{figure}
\centering
\includegraphics[width=0.45\textwidth]{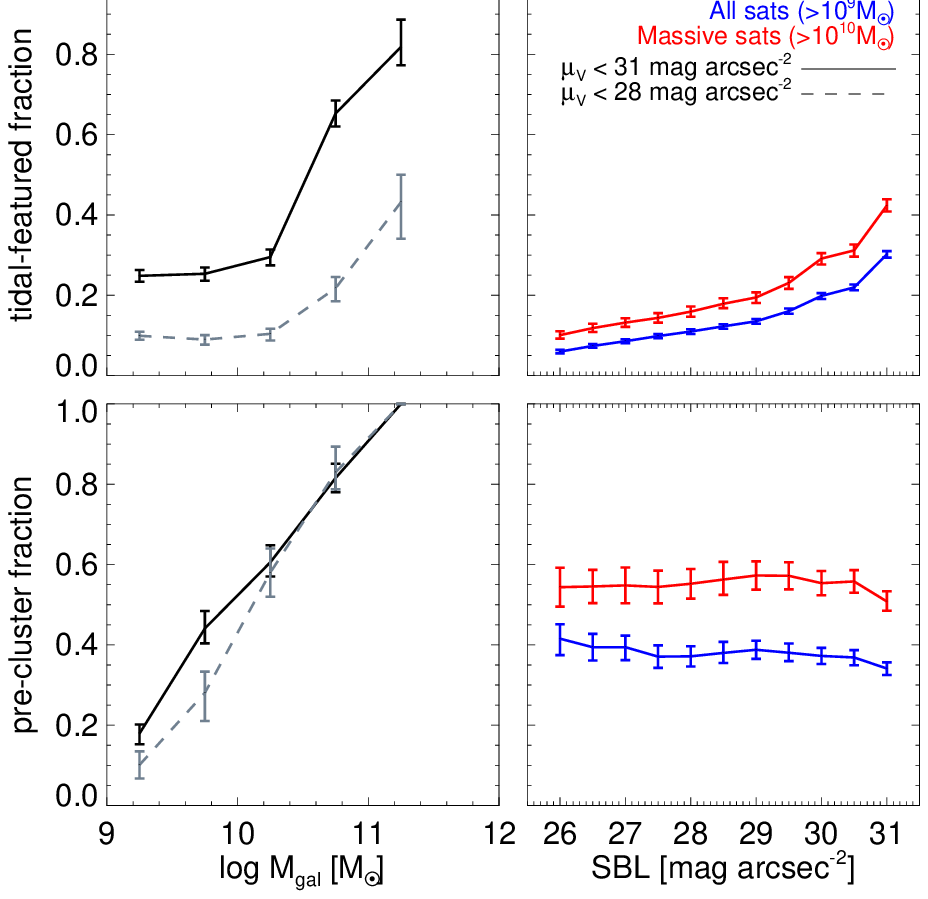}
\caption{The fractional distribution of tidal-featured galaxies in the cluster. Each panel shows the tidal-featured fraction (the top panels) and the pre-cluster fraction (the bottom panels), depending on the galaxy mass (the left panels) and V-band surface brightness limits (the right panels). The type and color of lines are the same as Figure \ref{fig:rprof}. The error bars represent the 1$\sigma$ uncertainties derived from bootstrap sampling. The bins containing fewer than five galaxies are omitted.}
\label{fig:mprof}
\end{figure}

In this subsection, we see how the surface brightness and the galaxy's lower-mass limits can affect the fractional distribution of the tidal-featured galaxies. For this, we generate the surface brightness map of tidal-featured galaxies, utilizing the projected 2D distribution of stellar particles within the $2\times R_{200c}$ of each galaxy. The stellar particles are projected onto three different planes: the x-y, x-z, and y-z planes and binned into a grid of cells with a length of D$\sim$1 ckpc\footnote{This length corresponds to a pixel scale of $\sim 2\arcsec$ at a distance of 100 Mpc.
This particular pixel specification is the same as that of the K-DRIFT (KASI-Deep Rolling Imaging Fast Telescope) pathfinder, which is a new telescope optimized for LSB studies \citep[][Ko et al. 2025 in preparation]{byun2022}.}. Due to the projection, the number of samples is tripled compared to the original number of tidal-featured galaxies. However, we normalize it to the original samples when showing the number of tidal-featured galaxies.

%Figure \ref{fig:rprof} shows the projected radial profile of the number of tidal-featured galaxies, the tidal-featured fraction (ratio of the tidal-featured galaxies to the cluster galaxies), and the pre-cluster fraction (ratio of the tidal-featured galaxies formed outside the cluster to the tidal-featured galaxies)\footnote{The pre-cluster galaxies are tidal-featured galaxies that form the tidal-feature both outside and inside the cluster, but before passing the pericenter.} with 1$\sigma$ uncertainties derived from bootstrap sampling.
Figure \ref{fig:rprof} shows the projected radial profile of the number of tidal-featured galaxies, the tidal-featured fraction (ratio of the tidal-featured galaxies to the cluster galaxies), and the pre-cluster fraction (ratio of the pre-cluster galaxies to the tidal-featured galaxies) with 1$\sigma$ uncertainties derived from bootstrap sampling. Here, pre-cluster galaxies refer to tidal-featured galaxies that form their tidal features both outside and inside the cluster, but before passing their first pericentric passage.
In the left panel of Figure \ref{fig:rprof}, the number of the tidal-featured galaxies increases as the projected clustercentric radius $R_{\text{proj}}/R_{\text{200c}}$ decreases. The higher frequency of tidal-featured galaxies in the inner region of the cluster suggests that the formation of a significant number of the tidal features is triggered in the inner region, near the BCG. Indeed, 50\% of galaxies that formed their tidal feature inside the cluster form the feature within 0.4$R_{200c}$ of the cluster. However, this result is contrary to the observations showing the more frequent or consistent fraction of tidal-featured and/or post-merger galaxies in the outer region of the clusters \citep{adams2012,sheen2012,oh2018}. The difference stems from the fact that the identification of the tidal feature used in this study is based on the dynamics of stellar components, not on morphological features. 
Furthermore, the BCG and ICL can hide tidal features in the inner regions of clusters.
More detailed discussion on this issue can be found in Section \ref{discussion}.

In the case of the massive galaxies with stellar mass more than $10^{10} \msun$, the number of the tidal-featured galaxies significantly decreases compared to that of all galaxies (the left panel of Figure \ref{fig:rprof}). On the contrary, the fraction of the massive tidal-featured galaxies increases compared to all galaxies (the middle panel of Figure \ref{fig:rprof}). It is because more massive galaxies are more likely to be tidal-featured galaxies (see the left top panel of Figure \ref{fig:mprof}), which is consistent with the observations \citep{sheen2012}. 
The higher fraction of tidal-featured galaxies among massive ones is due to their higher possibility of experiencing preprocessing, which accelerates tidal feature formation before cluster entry.
When we confine the tidal-featured galaxies to those identified with a brighter tidal feature than 28~mag~arcsec$^{-2}$, the tidal-featured fraction significantly decreases compared to that of 31~mag~arcsec$^{-2}$.
On the contrary, the pre-cluster fraction with $\mu_v<$~28~mag~arcsec$^{-2}$ significantly increases (the right panel of Figure \ref{fig:rprof}). This increase is more prominent in the outer region. Indeed, in the case of the massive galaxies with stellar mass more than $10^{10} \msun$, the fraction of the pre-cluster galaxies increases up to 80\% at the outer region of $R_{\text{proj}}/R_{\text{200c}}>0.4$.

Figure \ref{fig:mprof} shows how the tidal-featured fraction and the pre-cluster fraction are correlated with both the galaxy's stellar mass (the left panels) and the SB limits (the right panels). The higher tidal-featured fraction for the massive galaxies with stellar mass more than $10^{10}\msun$ can be explained by the increase of the pre-cluster tidal-featured galaxies.

The tidal-featured fraction increases as the SB limit increases (see the upper panels of Figure \ref{fig:mprof}), but in the case of the pre-cluster galaxies only, the dependencies get weaker since the tidal-featured galaxies formed inside the cluster (the opposite concept to the pre-cluster galaxies) are biased to the lower mass, whose tidal feature is also less bright. Indeed, the pre-cluster fraction increases from 0.3 to 0.4 when the SB limit varies from 31 to 26~mag~arcsec$^{-2}$ (see the right bottom panel of Fig. \ref{fig:mprof}). For massive galaxies whose stellar mass exceeds $10^{10}\msun$, the change in the pre-cluster fraction due to the SB limit is negligible.

Our results suggest not only the significance of preprocessing in the formation of tidal features but also that current observations of tidal-featured galaxies may be more biased toward pre-cluster galaxies.

\section{Discussion}\label{discussion}

\begin{figure*}
\centering
\includegraphics[width=0.9\textwidth]{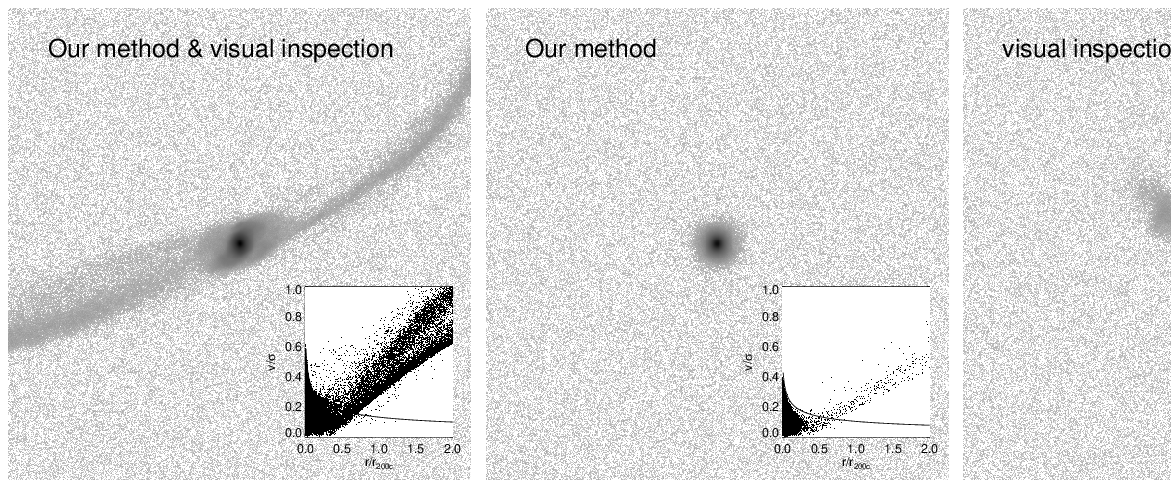}
\caption{Surface brightness maps (r$~<~$250~kpc) of tidal-featured galaxies identified by our method and visual inspection. From left to right, the columns show tidal-featured galaxies detected by both methods, by only our method, and by only visual inspection. The small panels show the phase-space diagram of the stellar particles of their infall members. The solid line indicates the escape velocities ($v_{\rm{esc}}$) derived using the current members, and the stars with velocities greater than $v_{\rm{esc}}$ are considered to be in the unbound regime.}
\label{fig:sbl_map}
\end{figure*}

\subsection{The definition of the tidal features}
As a result of Section \ref{result}, the $f_{\text{tide}}$ value is determined by a combination of several factors, such as 1) the $t_{\text{infall}}$ distribution, 2) the ratio of the preprocessed infallers that changes by the $t_{\text{infall}}$, 3) the $\log P$ of each infaller, which is a complex mix of the density distributions of the galaxy and the cluster, as well as the galactic orbit respect to the cluster center, and 4) the $\Delta t_{\text{pre}}$ that determines the degree of the preprocessing. Since the M$_{\text{200c}}$ and the $z_{\text{m50}}$ parameters are insufficient to represent these factors or their combination, the $f_{\text{tide}}$ distribution shows significant scatter even if the galaxy samples are restricted by observational limits, such as the galaxy's lower-mass limit or the SB limit.

To quantitatively identify the tidal feature of galaxies, in this study, we use the $v_{\text{esc}}$ as a function of the galactocentric radius $r$ and check whether the individual stars of the infall members exceed the $v_{\text{esc}}(r)$ (see the right panels of Figure \ref{fig:map}). Since $v_{\text{esc}}(r)$ is calculated from the mass distribution of star and DM particles of the current member, infall member stars exceeding $v_{\text{esc}}(r)$ indicate that they are indeed gravitationally unbound from the galaxy. Because the number of unbound stars is the basis for the identification of tidal features, our methodology has a limitation in that it can not detect tidal features formed by stars that are still bound. 

To understand how our methodology differs from visual inspection, we generate mock images of all galaxies in the clusters at $z=0$ and select the galaxies with the tidal features such as tidal tail, stream, and shell-like structure by the visual inspection.
We generate the mock images of galaxies in the same method as in Section \ref{limit}, and the luminosity of stellar particles is additionally smoothed by a Gaussian kernel. The smoothing kernel size is set as the distance to the 5th nearest neighbor star particle, using the software from \cite{martin2022}.
Finally, we add Gaussian background noise with a surface brightness limit of $\mu_V = $31~mag~arcsec$^{-2}$ to the map of each galaxy.
When mapping, we exclude contamination from the bright members of the BCG, ICL, and other galaxies. 
This ensures a fair comparison with the tidal features identified by our method, which also excludes these components.
We show example surface brightness maps of tidal-featured galaxies detected by our method and visual inspection and phase-space diagrams of their infall members in Figure \ref{fig:sbl_map}.

Among the 3,262 cluster galaxies, we find that 674 are classified as tidal-featured galaxies by visual inspection, and the median $f_{\rm{tide}}$ values for the 84 GRT clusters are from 0.19 to 0.21 within the three host mass bins of $13.6 < \log $M$_{\rm{200c}} < 14$, $14.0 < \log $M$_{\rm{200c}} < 14.4$, and $14.4 < \log $M$_{\rm{200c}} < 14.8$ at $z=0$.
We find no significant dependency on the fraction of tidal-featured galaxies determined by visual inspection of the cluster mass.
However, $f_{\rm{tide}}$ is lower than that determined by our methodology (see Figure \ref{fig:ftide}), and $f_{\rm{tide}}$ of relaxed clusters is lower than that of unrelaxed clusters.

We find that 134 tidal-featured galaxies selected by the visual inspection are not classified as tidal-featured galaxies in our methodology.
These galaxies fell into the cluster recently (median value of $t_{\rm{infall}} \sim 1.9~$Gyr) and thus are sufficiently massive (median value of M$_{\rm{gal}} \sim 2.1\times10^{10}\msun$ at $t_{\rm{infall}}$).
However, among them, preprocessed infallers have $1.3-4.2~$Gyr smaller $\Delta t_{\rm{pre}}$ than the tidal-featured galaxies classified by our methodology, and direct infallers lose a smaller amount of mass (median value of M$_{\rm{200c}}$/M$_{\rm{peak}} \sim 33\%$) until $z=0$ than the direct infallers (median value of M$_{\rm{tide}}$/M$_{\rm{peak}} \sim 16\%$; see Figure \ref{fig:mpeak}) classified by our methodology.
Therefore, they are not classified as tidal-featured galaxies by our methodology.
On the other hand, 445 of 985 tidal-featured galaxies classified by our methodology are not classified as tidal-featured galaxies by visual inspection.
We find that the tidal features of these galaxies are obscured by background noise, with a representative example shown in the middle panel of Figure \ref{fig:sbl_map}.
These galaxies are less massive (median value of M$_{\rm{gal}} \sim 4.8\times10^{9}\msun$ at t$_{\rm{infall}}$) and fall into the cluster early (median value of t$_{\rm{infall}} \sim 6.9~$Gyr).
Therefore, they prefer to form the tidal feature from the stripped main stellar body within the cluster but lose a small amount of stellar mass ($\sim$1.5\%) after falling into the cluster.
This stellar mass loss is sufficiently smaller than that of other 540 tidal-featured galaxies ($\sim$5.7\%) classified by both visual inspection and our methodology.
We find that these visually undetected tidal-featured galaxies are biased to the direct infallers ($\sim$ 75\%).
This difference leads to a discrepancy in the classification of tidal-featured galaxies between our method and visual inspection. 

\begin{figure}
\centering
\includegraphics[width=0.45\textwidth]{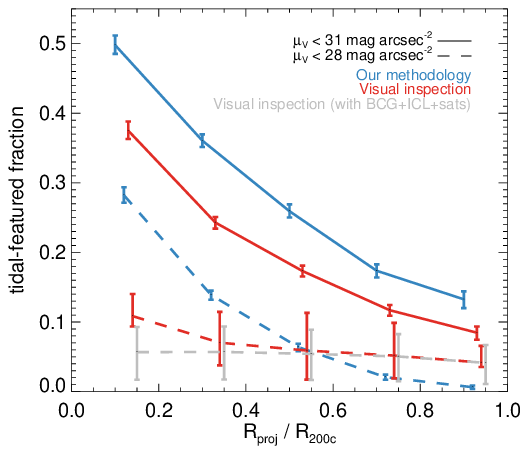}
\caption{The radial profile of tidal-featured fractions in the cluster. The solid and dashed lines indicate the tidal-featured fractions from brighter galaxies than 31~mag~arcsec$^{-2}$ and 28~mag~arcsec$^{-2}$. The fractions are measured using our method (the blue line), visual inspection (the red line), and visual inspection with the stellar components of the BCG, ICL, and satellites (the gray line). The error bars represent the 1$\sigma$ uncertainties derived from bootstrap sampling.}
\label{fig:vis}
\end{figure}

Figure \ref{fig:vis} highlights the difference in the tidal-featured fraction classified by both methods as a function of projected clustercentric distance.
The visually detected tidal-featured fraction with $\mu_V~<$~31~mag~arcsec$^{-2}$ (the solid lines) is lower than that identified from our methodology because a significant fraction (45\%) of the tidal-featured galaxies classified by our method is not detected by visual inspection, as mentioned above.
However, unlike previous observations \citep[e.g.,][]{adams2012,sheen2012,oh2018}, the visually detected fraction shows a radial trend similar to our methodology.

When we confine the analysis to brighter tidal features ($\mu_V~<~$28mag~arcsec$^{-2}$), the tidal-featured fraction between the two methods shows a more significant difference (dashed lines). Both methods still indicate an increase in the tidal-featured fraction with decreasing projected clustercentric radius, but visual inspection shows fewer tidal-featured galaxies in the inner region compared to our methodology. As mentioned above, tidal-featured galaxies detected only by our methodology are typically less massive with fainter tidal features and are more likely to reside in the cluster center due to their earlier infall \citep[see][]{rhee2017}.
This contributes to the lower visually detected tidal-featured fraction in the cluster center.
In contrast, in the outer region (R$_{\rm{proj}}/$R$_{\rm{200c}}~>~0.6$), the tidal-featured fraction classified by visual inspection is higher than that from our methodology, as visual inspection is more likely to detect tidal features of satellites that recently formed them, as mentioned above.

Although the visually detected tidal-featured fraction slightly increases toward the cluster center (increasing from 5\% to 10\%), it is essential to consider that the presence of the BCG and ICL in the inner regions of clusters can obscure tidal features, further reducing the visually detected tidal-featured fraction. This reduction might lead to a consistent or even decreasing fraction toward the cluster center, consistent with trends observed in previous studies \citep{adams2012,sheen2012,oh2018}.
Indeed, we find that the tidal-featured fraction ($\mu_V~<~$28mag~arcsec$^{-2}$) measured by visual inspection using the maps contaminated by the stellar components of the BCG, ICL, and satellites shows a constant radial profile (the gray dashed line).
Regardless, our results show that many direct infallers in the central region hide their tidal features, indicating that the observed tidal-featured galaxies may be more biased toward preprocessed infallers.

\subsection{The formation of ultra-diffuse galaxies in the cluster}
Many observational studies have discovered a lot of ultra-diffuse galaxies (UDGs) within galaxy groups and clusters \citep[e.g.,][]{koda2015,mihos2015,vdokkum2015,iodice2020,lim2020}.
They are commonly defined as the low-mass galaxies ($M_{\rm{gal}} < 10^9\msun$) with the low central surface brightness ($\mu_{g,0} >$ 24.0 mag/arcsec$^2$) and with large half light radius ($R_{\rm{e}} >$ 1.5~kpc) \citep[e.g.,][]{vdokkum2015,gannon2024}.
Their formation scenarios are various, but in dense environments such as galaxy groups and clusters, the normal satellites can also evolve into UDGs as they undergo tidal stripping and heating \citep[e.g.,][]{ogiya2018,sales2020,tremmel2020}.
These UDGs can have tidal features, and a recent observational study has indeed shown the presence of UDGs with tidal features \citep[][]{zemaitis2023}.

In our simulation, we find that the 977 satellites\footnote{We only analyze the low-mass galaxies with M$_{\rm{gal}} < 10^9\msun$ in this section because their populations are incomplete as mentioned in Section \ref{method}.} show the characteristics of UDGs.
Among them, 472 UDGs have the tidal features determined by our methodology at $z=0$. 
We find that these UDGs fall into the cluster earlier (median value of $t_{\rm{infall}} \sim $7.7~Gyr ago) than other UDGs (median value of $t_{\rm{infall}} \sim $5.6~Gyr ago).
Therefore, the UDGs with tidal features lose their mass more significantly (median value of M$_{200c}/$M$_{\rm{peak}} \sim $11\% at $z=0$) and more concentrated to the central region of clusters (median value of R$_{proj}/$R$_{200c} \sim $0.36) than other UDGs (median value of M$_{200c}/$M$_{\rm{peak}}$ and R$_{proj}/$R$_{200c} \sim $39\% and 0.5 at $z=0$).
However, we cannot find evidence that the two UDG groups are significantly distinguished by the classifications of the preprocessed/direct infallers and/or the pre-cluster/in-cluster galaxies.
Although many UDGs have tidal features determined by our methodology, their tidal features are usually undiscovered in the observation.
We find that 14\% (66) of UDGs with tidal features are classified as tidal-featured galaxies by visual inspection of 31~mag~arcsec$^{-2}$ and only two UDGs show their tidal features brighter than 28~mag~arcsec$^{-2}$.
This indicates that upcoming deeper imaging observations will reveal evidence of tidal stripping of UDGs, such as the recently observed tidal tail near F8D1 UDG in the M81 group \citep{zemaitis2023}.

\begin{figure}
\centering
\includegraphics[width=0.45\textwidth]{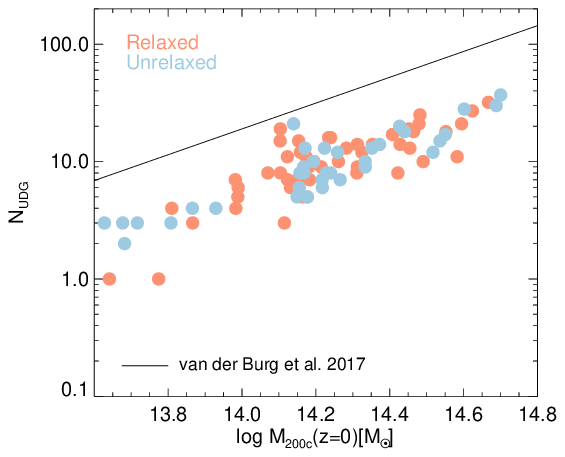}
\caption{The abundance of UDGs in the GRT clusters. The filled red and blue circles indicate the relaxed and unrelaxed clusters. The black solid line is the best-fitting power-law relation (N$_{\rm{UDG}} \propto$ M$_{\rm{200c}}^{1.11}$) between the abundance of UDGs and the host mass from \cite{vburg2017}.}
\label{fig:udg}
\end{figure}

We additionally find that 578 satellites were transformed into UDGs within the cluster environment, while others were already classified as UDGs at their infall time.
In the GRT scheme, the replaced high-resolution stellar galaxy is modeled as a bulgeless exponential disk galaxy, and the mass of the stellar disk is determined using the stellar-to-halo mass relation as a function of the redshift from \cite{behroozi2013a}. 
Moreover, since the in situ star formation is not considered after the high-resolution model is inserted into the DM halo, UDGs cannot form in the field or through hydrodynamic recipes in the GRT simulations.
Therefore, UDGs identified as such before cluster infall must have been transformed into UDGs by tidal interaction outside the cluster.
Indeed, they already undergo a significant mass loss before falling into the cluster, with a median $M_{\rm{200c}}/M_{\rm{peak}} \sim 64\%$ at t$_{\rm{infall}}$.
As a result, in our simulations, UDGs might be formed only by tidal stripping, resulting in fewer UDGs than the observed UDGs in the clusters (Figure \ref{fig:udg}).
Figure \ref{fig:udg} shows the number of UDGs in 84 individual GRT clusters.
The black line is the best-fitting power-law relation between the abundance of UDGs and the host mass from \cite{vburg2017}, and the filled red and blue circles indicate the number of UDGs within relaxed and unrelaxed clusters in our simulations.
In this figure, we can see the abundance of UDGs is not related to the dynamical states of the clusters.
Moreover, the simulated clusters have an average of 28\% of the UDGs compared to the observed clusters, regardless of host mass.
These results suggest that tidal stripping, both inside and outside the cluster, is not the dominant formation mechanism for UDG formation in the clusters.

\subsection{The formation of shell-like galaxies}
Observation and simulation studies have suggested that the shell-like galaxies are formed through mergers with galaxies of comparable mass or with smaller companions on radial orbits \citep[i.e.,][]{hernquist1988,ebrova2012,paudel2017,pop2018,karademir2019,yoon2024}. These galaxies have been observed in both group and cluster environments \citep[i.e.,][]{malin1983,paudel2017,lima2020}, but it remains unclear whether their formation primarily occurs inside or outside group or cluster environments. To investigate this, we trace whether shell-like galaxies are classified as preprocessed or direct infallers and pre-cluster or in-cluster galaxies.

\begin{figure}
\centering
\includegraphics[width=0.45\textwidth]{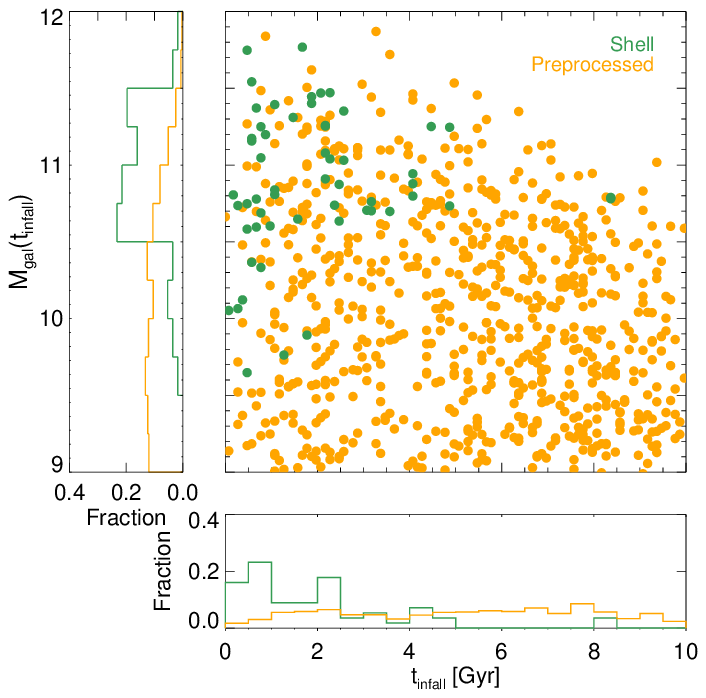}
\caption{The galaxy mass of preprocessed infallers with tidal features when they fall into the cluster. The filled green and yellow circles indicate the shell-like galaxies and other preprocessed infallers with tidal features. The left and bottom histograms indicate the distribution of M$_{\text{gal}}$($t_{\text{infall}})$ and $t_{\text{infall}}$.}
\label{fig:shell}
\end{figure}

In total, 57 shell-like galaxies are identified by visual inspection, and all but one belong to the preprocessed infallers. The percentage of the shell-like galaxies is $\sim$7\% of the preprocessed infallers with tidal features. Moreover, all shell-like galaxies belonging to the preprocessed infallers are the pre-cluster galaxies that form the shell feature before the cluster entry (81\%) or before the first pericenter (19\%). The percentage of the shell-like galaxies is 16\% of the pre-cluster galaxies. Figure \ref{fig:shell} shows distributions of the stellar mass (M$_{\text{gal}}$) of the preprocessed infallers with tidal features and the shell-like galaxies at their infall time. As shown in previous studies \citep[e.g.,][]{pop2018,yoon2024}, the shell-like galaxies are relatively more massive than the other galaxies; M$_{\text{gal}}$($t_{\text{infall}})$ of the shell-like galaxies peaks at $6.9\times10^{10}\msun$, while that of the other preprocessed infallers with tidal features peaks at $9.4\times10^{9}\msun$. The median $t_{\text{infall}}$ value is also a bit different; the shell-like galaxies entered the cluster more recently, at 1.7~Gyr ago, while the other preprocessed infallers with tidal features entered at 5.6~Gyr ago. Moreover, all shell-like galaxies classified as preprocessed infallers had host satellites before they entered the cluster, compared to only 37\% of the other preprocessed infallers with tidal features. 
These findings suggest that the shell-like galaxy in the cluster prefers to be formed by interaction with its satellites before falling into the cluster.

\begin{figure}
\centering
\includegraphics[width=0.45\textwidth]{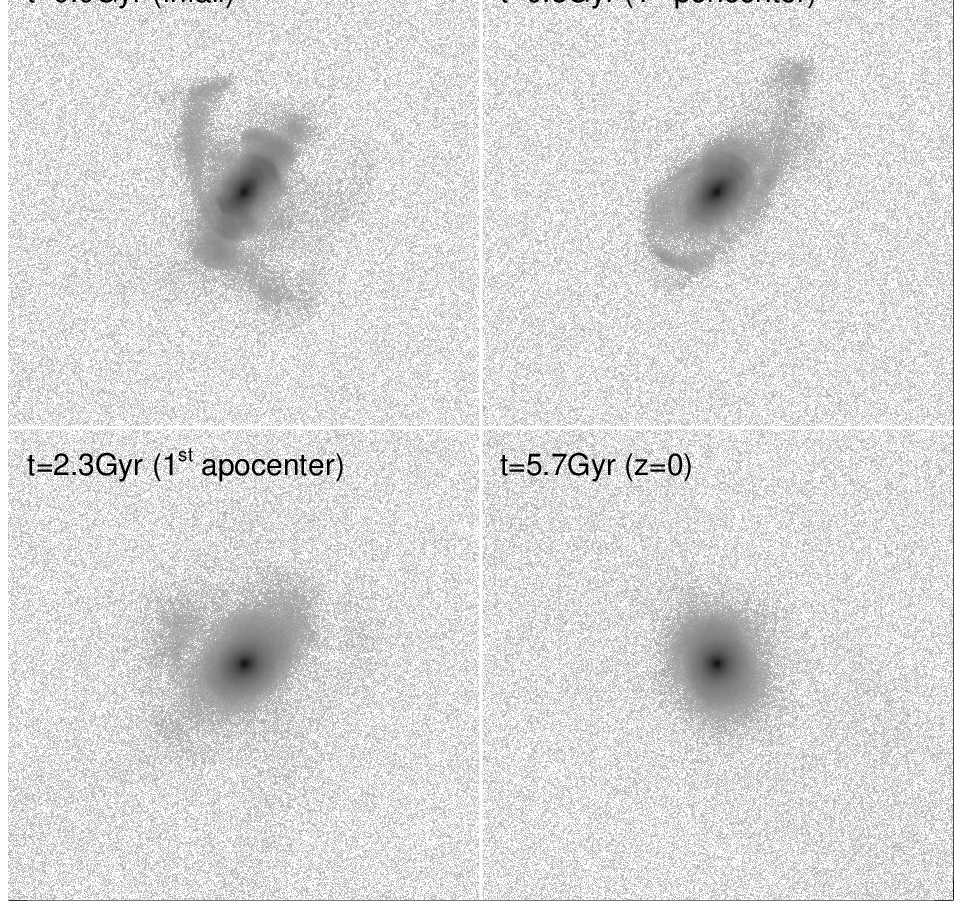}
\caption{The evolution of the example shell-like galaxy. From top left to bottom right, panels show the stellar distribution at the time of cluster infall, first pericenter passage, first apocenter passage, and $z=0$. The time (t) in the top left corner of each panel indicates the elapsed time since infall.
}
\label{fig:shell_image}
\end{figure}

We additionally find that eight preprocessed infallers had shell-like features when they fell into the cluster, but these structures do not survive until $z=0$. These galaxies had fainter shell-like features, as they fell into the cluster earlier ($t_{\text{infall}}~\sim~$4.7~Gyr) with lower mass (M$_{\text{gal}}~\sim~3.5\times10^{10}\msun$) than the shell-like galaxies at $z=0$.
The time these galaxies have spent in the cluster exceeds the expected survival time of shell-like features \citep[$\sim$4 Gyr;][]{mancillas2019}. Moreover, as satellite galaxies experience the strong tidal field of the cluster, tidal stirring by the cluster potential randomizes the orbits of stars and erases faint substructures \citep{lokas2020}. Thus, shell-like features are rapidly erased, and the galaxies are transformed morphologically. Figure \ref{fig:shell_image} illustrates the evolution of the example shell-like galaxy: the galaxy initially exhibited shell-like features, but these had largely faded by the time it reached its first apocenter, and it eventually evolves into a spheroidal system by z=0.
%Although we do not fully understand why their shell-like features are invisible at $z=0$, the strong tidal field in the cluster can destroy the features more quickly than the isolated environment.
%Moreover, the time the galaxies have spent in the cluster is longer than the survival time of the shell-like feature expected in the simulation \citep[4~Gyr;][]{mancillas2019}.

\section{Summary}
\label{sec:summary}

In this study, we investigate the tidal features of the satellites of M$_{\rm{gal}} > 10^9 \msun$ in the 84 clusters of $13.6 < \log$ M$_{\rm{200c}} < 14.4$ at $z=0.0$ using cosmological $N-$body simulations.
To trace the formation and evolution of the stellar structures in the cosmological $N-$body simulations, we use an alternative simulation technique referred to as the ``galaxy replacement technique" (GRT), first introduced in \cite{chun2022}.
Because the GRT does not include computationally expensive baryonic physics, we can accurately trace the spatial distribution and evolution of tidal features near the galaxies with sufficient high-resolution star particles.

To identify the tidal features of satellites in the clusters, we trace all member particles of each galaxy after the moment when they first enter the clusters and consider the members with velocities greater than escape velocity ($v_{esc}$) at a given distance from the galaxy center to be in an unbound regime.
We classify satellites as tidal-featured galaxies if they contain more than $10^7 \msun$ of stellar mass in the unbound regime and as non-tidal-featured galaxies if their stellar mass in the unbound regime is below $10^6 \msun$.

In this study, we aim to understand how satellites can have tidal features in the cluster environment and the effect of preprocessing on the formation of tidal features.
Moreover, we investigate the spatial and mass distributions of the satellites with tidal features and the effect of the detection limits on observing the tidal features near the satellites.
We summarize our results as follows:

\begin{enumerate}\setlength{\itemsep}{-1mm}
\item The fraction of satellites with tidal features ($f_{\rm{tide}}$) increases with the mass of the host clusters but shows no significant dependence on the dynamical state of clusters (Figure \ref{fig:ftide}). This mass dependency weakens when tidal-featured galaxies are classified via visual inspection.
\item For preprocessed infallers, the tidal features correlate more directly with the degree of preprocessing outside the cluster than with tidal perturbation inside the cluster (Figure \ref{fig:logp}-\ref{fig:pre}). We find that a significant fraction (43\%) of preprocessed infallers with tidal features formed the tidal features even before entering the cluster or before the first pericenter.
\item Direct infallers with tidal features formed their tidal features within the cluster environment. They typically fell into the cluster earlier (median value of $t_{\rm{infall}} \sim $8.5~Gyr ago, and thus significantly lose their mass in the cluster (M$_{\rm{200c}}/$M$_{\rm{peak}} \sim 16\%$ at $t_{\rm{tide}}$) due to the strong tidal perturbation in the cluster environment (See Figure \ref{fig:type_infall}-\ref{fig:logp}).
\item 49\% of satellites in our cluster samples experienced preprocessing before infall, whereas the fraction of preprocessed infallers among the tidal-featured galaxies is higher (79\%). Thus, preprocessing boosts the formation of tidal features and is the dominant path for forming the tidal features of the satellites in the cluster.
\item Many less massive satellites form tidal features in the central region of clusters, but are usually missed by visual inspection. On the other hand, the fraction of pre-cluster galaxies increases with radius. This increase is more emphasized in massive galaxies and brighter surface brightness limits. This indicates that the current observations may be more biased toward pre-cluster galaxies (See Figure \ref{fig:rprof}-\ref{fig:mprof}).
\end{enumerate}

These results are derived using stellar components brighter than $\mu_{\rm{V}} < $ 31~mag~arcsec$^{-2}$, corresponding to the resolution limit of the GRT simulation. Moreover, the tidal features are identified by the internal dynamics of the galaxies.
Although a different detection method (i.e., visual inspection) may change the identified population of satellites with tidal features, we find that the pre-cluster galaxies are still the dominant population (68\%).
Our results show that the preprocessing significantly affects the formation of tidal features such as the tidal tail, stream, and shell-like structures in the cluster environment.
Moreover, many cluster galaxies will reveal their tidal features through upcoming deeper imaging observations, which will provide us more information about their formation and evolution.

\bibliography{main}{}
\bibliographystyle{aasjournal}

\end{document}